\documentclass[aps, prb, reprint, showpacs, twocolumn, 10pt, superscriptaddress]{revtex4-1}

\usepackage{graphicx, amsmath, dsfont, dcolumn, bm, times, color, braket, ulem, bbold}

\usepackage[usenames,dvipsnames]{xcolor}

\definecolor{mygreen}{rgb}{0.0, 0.5, 0.0}

\begin{document}

\bibliographystyle{apsrev4-1}

\title{Voltage induced conversion of helical to uniform nuclear spin polarization in a quantum wire}

\author{Viktoriia Kornich}
\affiliation{Department of Physics, University of Basel, Klingelbergstrasse 82, CH-4056 Basel, Switzerland}
\author{Peter Stano}
\affiliation{RIKEN Center for Emergent Matter Science, 2-1 Hirosawa, Wako, Saitama 351-0198, Japan}
\author{Alexander A. Zyuzin}
\affiliation{Department of Physics, University of Basel, Klingelbergstrasse 82, CH-4056 Basel, Switzerland}
\author{Daniel Loss}
\affiliation{Department of Physics, University of Basel, Klingelbergstrasse 82, CH-4056 Basel, Switzerland}
\affiliation{RIKEN Center for Emergent Matter Science, 2-1 Hirosawa, Wako, Saitama 351-0198, Japan}

\date{\today}

\begin{abstract}
We study the effect of bias voltage on the nuclear spin polarization of a ballistic wire, which contains electrons and nuclei interacting via hyperfine interaction. In equilibrium, the localized nuclear spins are helically polarized due to the electron-mediated Ruderman-Kittel-Kasuya-Yosida (RKKY) interaction. Focusing here on non-equilibrium, we find that an applied bias voltage induces a uniform polarization, from both helically polarized and unpolarized spins available for spin flips. Once a macroscopic uniform polarization in the nuclei is established, the nuclear spin helix rotates with frequency proportional to the uniform polarization. The uniform nuclear spin polarization monotonically increases as a function of both voltage and temperature, reflecting a thermal activation behavior. Our predictions offer specific ways to test experimentally the presence of a nuclear spin helix polarization in semiconducting quantum wires.
\end{abstract}

\pacs{62.23.Hj, 75.75.-c, 73.21.-b, 31.30.Gs }

\maketitle

\let\oldvec\vec
\renewcommand{\vec}[1]{\ensuremath{\boldsymbol{#1}}}

\section{Introduction}
\label{sec:Introduction}
Magnetic structures are promising platforms for many modern devices, e.g. memory,\cite{parkin:science08} sensors,\cite{rugar:naturenano15} and quantum computation hardware. \cite{kloeffel:arcmp13}The opportunities to get an ordered magnetic  phase in the bulk and low-dimensional systems due to Ruderman-Kittel-Kasuya-Yosida (RKKY) interaction \cite{froehlich:prsl40, ruderman:pr54, kasuya:ptp56, yosida:pr57} were studied in a number of theoretical and experimental works.\cite{dietl:prb97, ohno:science98, simon:prl07, simon:prb08, ohno:nature00, chiba:nature08, braunecker:prl13} The prominent feature of RKKY interaction in 1D systems is the ordering of localized spins into a helix.\cite{braunecker:prl09, braunecker:prb09} 

When the current is driven through the system of electrons and nuclei, the spin polarization can be swapped between the two subsystems through the hyperfine interaction, leading to dynamic nuclear polarization effects.\cite{dixon:prb97, machida:prb02, deviatov:prb04, trowbridge:arxiv, chida:prb12, chesi:arxiv15} If the polarization of current carrying electrons and localized spins differ, the spin-transfer torque arises,\cite{slonczewski:jmmm96, berger:prb96} important for dynamics of domain walls\cite{tatara:pr08, li:prl12} and enhancing the tilting of the spiral structure in helimagnets.\cite{hals:prb13} Closely related is the dynamic nuclear polarization, arising e.g. in helical edge states of topological insulator. The backscattering of helical electrons can be of different origins, such as assisted by phonons,\cite{budich:prl12} magnetic impurities,\cite{maciejko:prl09} or absence of axial spin symmetry.\cite{schmidt:prl12} It was shown that nuclear-assisted backscattering of electrons due to hyperfine interaction induces nuclear polarization when the current is driven through the edge states of topological insulator.\cite{lunde:prb12, delmaestro:prb13}

The main motivation for our work comes from the recent experiment by Scheller {\it et al.},\cite{scheller:prl14} where the conductance of a cleaved edge overgrowth GaAs quantum wire was measured. The measurements showed that the conductance of the first mode becomes $e^2/h$ at low temperatures instead of the naively expected $2e^2/h$. This suggests the lifting of electron spin degeneracy. The possible explanation is the presence of a  helical nuclear spin polarization that gaps out one subband and thus provides an electron spin selection. Further ways to confirm the  presence of the nuclear spin helix were suggested theoretically, by means of nuclear magnetic resonance,\cite{stano:prb14} nuclear spin relaxation,\cite{zyuzin:prb14} and quantum Hall effect anisotropies.\cite{meng:arxiv14} 

In this work we propose and study a complementary method to detect nuclear spin helical polarization in the wire. It is based on the effect of bias voltage applied to the wire and therefore straightforward to perform experimentally. We investigate how the bias voltage applied to the wire affects its nuclear spin polarization. We assume that at zero bias and finite temperature, nuclear spins are partially polarized into a helix due to the RKKY interaction. We find that an applied voltage induces a uniform nuclear polarization from both helical and non-polarized nuclear spins available for nuclear spin flips via electrons. Therefore, upon increasing the voltage the helical nuclear polarization drops, while the uniform polarization grows, and the total polarization grows too. For small voltages and increasing temperature,  
the uniform polarization grows because of thermal activation of electrons, while the helical polarization dramatically drops in magnitude.
Once a macroscopic uniform polarization has developed, the remaining nuclear spin helix rotates as a whole around the axis along the uniform polarization. 
Since the helical polarization affects the conductance of such systems,~\cite{braunecker:prl09, braunecker:prb09,scheller:prl14,meng:arxiv14}
these predicted features are expected to show up in the voltage and temperature dependence of the transport currrent and thus they can be tested experimentally.  
Recently,  cantilever-based magnetic sensing techniques have been reported which enable nuclear spin magnetometry of nanoscale objects such as the nanowires considered here.\cite{peddibhotla:nature13} Such powerful techniques offer promising perspectives for direct experimental tests of the results obtained in this work.

The paper is organized as follows. In Sec. \ref{sec:TheModel} we present the Hamiltonian of our model. In Sec. \ref{sec:BathAsOneDimensionalChiralElectrons} we describe the properties of the electron bath. The derivation of the Bloch equation for the total nuclear spin in the wire is discussed in Sec. \ref{sec:BlochEquation}. The resulting nuclear spin polarization and its dependence on the parameters of the system are presented and discussed in Sec. \ref{sec:Polarization}. Our conclusions follow in Sec. \ref{sec:Conclusions}. Additional information about our calculation is given in Appendix.   
    
\section{The model}
\label{sec:TheModel}
We consider a one-dimensional electron gas and localized spins in a semiconductor nanowire. We will refer to these localized spins as nuclear spin in the following, however, they can be also of other origins, such as e.g. magnetic impurities, etc. The electrons and nuclei interact via the hyperfine interaction described by the Hamiltonian
\begin{equation}
H_{hyp}=\frac{1}{2}A\rho_0^{-1}|\psi_\perp(\bm{R}_\perp)|^2\delta(r-R)\bm{\sigma}\cdot\bm{I},
\label{eq:HyperfineHamiltonian}
\end{equation}
where $A$ is a hyperfine constant of the material, $\rho_0$ is the nuclear spin density, $\psi_\perp$ is the transverse part of electron wavefunction, $r$ denotes the electron position along the wire, $(R, \bm{R}_\perp)$ is the position of the nucleus along the wire and in the transverse direction respectively, $\bm{\sigma}$ is an electron spin operator, and $\bm{I}$ is a nuclear spin operator (in units of $\hbar$) with the magnitude $I$. We assume that 
the transverse part of the electron wavefunction $\psi_\perp(\bm{R}_\perp)$ is constant in the wire cross-section, 
$\ |\psi_\perp(\bm{R}_\perp)|^2=1/C$, where $C$ is the wire crossection area.
We parametrize it alternatively by the number of nuclear spins in the cross-section, $N_\perp=C a \rho_0$, with $a$ being the lattice constant. In GaAs $\rho_0=8/a^3$, $a=0.565$ nm, $A=90\ \mu\mbox{eV}$, $I=3/2$, and $N_\perp$ is typically of the order of $10^3$. Finally, we introduce $N=L/a$ with $L$ the wire length (typically of order microns), which gives $N N_\perp$ as the total number of nuclear spins in the wire.

The total Hamiltonian reads
\begin{equation}
H_{tot}=-\frac{\hbar^2}{2m}\partial_r^2+H_{hyp},
\end{equation}
where $m$ is electron effective mass and $\hbar$ is the Planck constant.
If the hyperfine interaction, Eq. (\ref{eq:HyperfineHamiltonian}), is weak on the energy scale of the electrons, its effects can be treated perturbatively. The condition is quantified by $A\ll\varepsilon_F$, where $\varepsilon_F$ is the Fermi energy of the electron system. 
This condition is well satisfied in the cases we consider here. A Schrieffer-Wolff transformation on $H_{tot}$ perturbatively in $H_{hyp}$, {\it i.e.},  in order $A/\varepsilon_F$, results to leading order in an effective interaction between the localized spins, the RKKY interaction,\cite{froehlich:prsl40, ruderman:pr54, kasuya:ptp56, yosida:pr57, simon:prb07, simon:prb08} 
\begin{equation}
\label{eq:RKKY}
H_{\scriptscriptstyle{RKKY}}=\sum_{i,j}\bm{I}_i \cdot J_{ij}  \bm{I}_j.  
\end{equation}
Here, the indexes $i,j$ label the nuclear spins and the RKKY coupling $J_{ij}=J(|R_i-R_j|)$ is related to the static spin susceptibility of electrons (see Eq.~(C1) and below in Ref.~\onlinecite{meng:arxiv14}), giving rise to the spatially dependent RKKY interaction.

\begin{figure}[tb]
\begin{center}
\includegraphics[width=\linewidth]{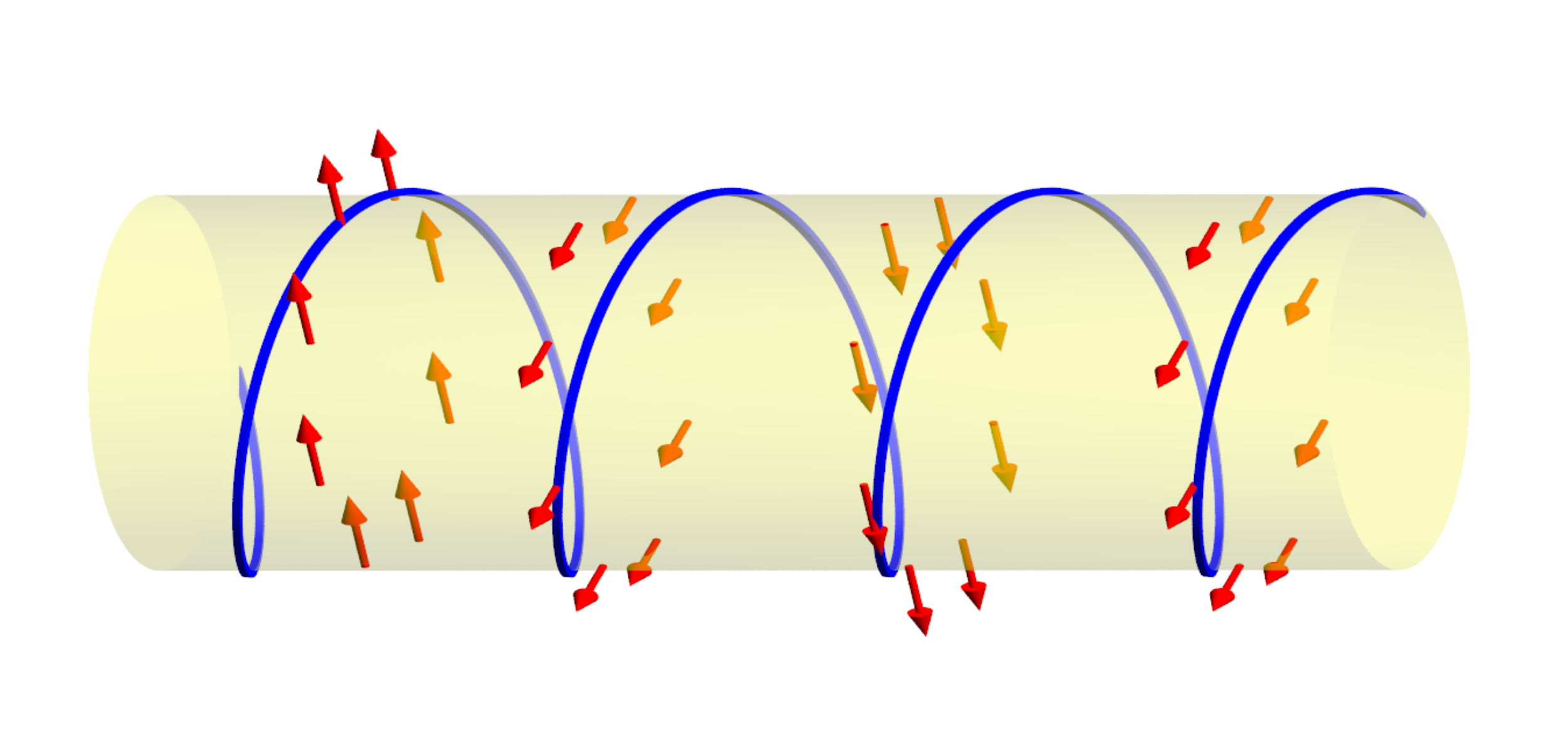}
\caption{A sketch of a conducting wire (yellow cylinder) with itinerant electrons (not shown) that couple to localized nuclear spins (red arrows) via hyperfine interaction. As a result,  a helical nuclear polarization emerges below a critical temperature. The blue spiral is a guide to the eye showing the direction of the helical polarization. The helical plane is chosen to be perpendicular to the wire axis (which need not be the case in general).}
\label{fig:nanowire}
\end{center}
\end{figure}

Let us rewrite Eq.~\eqref{eq:RKKY} in the momentum representation, defined through the Fourier transforms $J_q=\sum_{R_i} \exp[-i q (R_i-R_j)] J_{ij}$, with $R_i \in a, 2a, \ldots, Na$, and ${\bf I}_q=\sum_i \exp(i q R_i) {\bf I}_i$, with $i\in 1, \ldots, N N_\perp$, and in both cases $q \in (2\pi/N) \times \{ 0,1,\ldots, N-1 \}$. We get
\begin{equation}
H_{\scriptscriptstyle{RKKY}}= \frac{1}{N}\sum_q {\bf I}_q \cdot J_q {\bf I}_{-q}.
\label{eq:RKKY in q space}
\end{equation}
In one dimension, the RKKY coupling $J_{q}$ has a sharp minimum at momentum $q=\pm 2k_F$, with $k_F=\sqrt{2m\varepsilon_F}/\hbar$ the electron Fermi wavevector.\cite{braunecker:prl09, braunecker:prb09} Consider an approximation in which we neglect all values of $J_q$ with respect to the large (negative) value at this minimum,
\begin{equation}
H_{\scriptscriptstyle{RKKY}} \simeq \frac{1}{N} J_{2k_F} \left( {\bf I}_{2k_F} \cdot {\bf I}_{-2k_F} + {\bf I}_{-2k_F} \cdot {\bf I}_{2k_F} \right).
\label{eq:RKKY in q space simplified}
\end{equation}
To understand the spectrum of this Hamiltonian, we introduce linearly transformed spin operators, 
\begin{equation}
\bm{I}_i=\mathcal{R}_{\bm{u},2k_FR_i} \bm{\hat{I}}_i,
\label{eq:rotated frame}
\end{equation}
with $\mathcal{R}_{{\bf u},\phi}$ the matrix corresponding to a rotation by angle $\phi$ around a unit vector ${\bf u}$. Inserting Eq.~\eqref{eq:rotated frame} into Eq.~\eqref{eq:RKKY in q space simplified} we get
\begin{equation}
H_{\scriptscriptstyle{RKKY}} \simeq \frac{1}{N} J_{2k_F} \left( \bm{\hat{I}}_{q=0}^{\perp} \cdot \bm{\hat{I}}_{q=0}^{\perp}  + \mathcal{B}/2 \right),
\label{eq:ferromagnet}
\end{equation}
where we define the vector components along $\bm{u}$ as $\hat{I}^u=\bm{\hat{I}}\cdot\bm{u}$, and perpendicular to it as $\bm{\hat{I}}^\perp = \bm{\hat{I}}-\hat{I}^u\bm{u}$, and we separated 
the terms bilinear in the spin operators at finite momenta,
\begin{equation}
\mathcal{B} = \sum_{q=\pm 2k_F} [2 \hat{I}_{q}^u \hat{I}_{-q}^u +   
\bm{\hat{I}}_{2q}^{\perp} \cdot \bm{\hat{I}}_{-2q}^{\perp} +
i {\rm sgn}(q)(\bm{\hat{I}}_{2q}^{\perp} \times \bm{\hat{I}}_{-2q}^{\perp} )  \cdot \bm{u}].
\label{eq:ugly B}
\end{equation}

The first term in the bracket of Eq.~\eqref{eq:ferromagnet} describes the energy of ferromagnetically coupled spins $\hat{\bm{I}}_i$: a configuration in which all these spins are collinear, along a vector perpendicular to $\bm{u}$, gives a minimal possible energy, of value $N N_\perp^2 J_{2k_F} I^2$. This configuration corresponds to a classical ground state of Eq.~\eqref{eq:RKKY in q space} as well, as it saturates the energy lower bound obtained using the Parseval's identity $\sum_q |\bm{I}_q|^2 =N N_\perp \sum_i |\bm{I}_i|^2$. Going back to the laboratory frame according to Eq.~\eqref{eq:rotated frame}, the ground state corresponds to a helical configuration where the nuclear spins are oriented parallel to each other in the wire cross-section, along a direction which rotates in a fixed plane as one moves along the wire (for illustration, see Fig. \ref{fig:nanowire}). We shall refer to this plane as the helical plane, with $\bm{u}$ being its normal unit vector. A unit vector $\bm{h}\perp \bm{u}$ gives the direction of the polarization within this plane at position $R=0$.

The finite momenta components, Eq.~\eqref{eq:ugly B}, arise from the choices of a definite helicity and the vector $\bm{u}$ in Eq.~\eqref{eq:rotated frame}, which break the full spin rotational symmetry of Eq.~\eqref{eq:RKKY in q space simplified}. 
Namely, choosing a frame with helicity opposite to the ground state helicity would lead to a swap of the roles of $\hat{\bm{I}}^\perp_0$ and $\hat{\bm{I}}^\perp_{\pm 4k_F}$. Further, configurations where both helicities are populated lead to a lower energy gain. For example, choosing both with the same weight, gives in the laboratory frame a spin-density wave, {i.e.}, a cos-like oscillation along a fixed vector, $\bm{I}_i =\bm{h} \cos{(2k_F R_i)} $, which gives only half of the energy gain of a helical order. Such oscillating, rather than rotating, configuration corresponds to the first term in Eq.~\eqref{eq:ugly B}. To conclude, up to the spin rotational symmetry, which allows for arbitrary directions of $\bm{u}$, and $\hat{\bm{I}}^\perp$, the ground state with ferromagnetically aligned $\hat{\bm{I}}_i$ (helically ordered $\bm{I}_i$) is unique.

If the order is established, the expectation value of $\hat{\bm{I}}^\perp_{q=0}$ is macroscopic, and we parametrize it by a polarization $p_h$,   
\begin{equation}
\langle \hat{\bm{I}}^\perp_{q=0} \rangle = N N_\perp I p_h \bm{h},
\end{equation}
so that $p_h=1$ corresponds to a completely ordered state. With this we reduce Eq.~\eqref{eq:RKKY in q space simplified} by the mean field approximation to a Hamiltonian describing a set of non-interacting spins
\begin{equation}
\label{eq:RKKYAveraged}
H_{\scriptscriptstyle{RKKY}}\simeq  \sum_i \mu_N \bm{B}_{i}^N\cdot \bm{I}_i,
\end{equation}
in the presence of the position-dependent internal field
\begin{equation}
\label{eq:omega0}
\mu_N \bm{B}_{i}^N = 2 p_h N_\perp I J_{2k_F} \mathcal{R}_{\bm{u}, 2k_FR_i} \bm{h}.
\end{equation}
This concludes a simplified derivation of the reduction of the RKKY Hamiltonian, Eq.~\eqref{eq:RKKY}, into a set of non-interacting spins, Eq. (\ref{eq:RKKYAveraged}), in an effective (mean) field, Eq. (\ref{eq:omega0}). A detailed analysis of the applicability of such an approximation was given in Ref.~\onlinecite{meng:arxiv14}, based on the derivation of the spectrum of the full Hamiltonian Eq.~\eqref{eq:RKKY}, without employing a mean field ansatz. There it was found that this approximation, in essence neglecting the long wavelength magnons, is well justified for sub-Kelvin temperatures and wire lengths relevant for mesoscopic experiments.

As we consider the limit $A\ll\varepsilon_F$, we adopt the Bohr-Oppenheimer approximation, assuming that electrons react instantaneously to the changes in nuclear spin subsystem. Consequently, we can consider the effect of the nuclear polarization on electrons as an Overhauser field\cite{braunecker:prb09} 
\begin{equation}
\label{eq:BOv}
\mu_e\bm{B}_{Ov}=\frac{Aa}{2N_\perp}\sum_j\delta(r-R_j)\langle\bm{I}_j\rangle,
\end{equation}
where $\mu_e$ is an electron magnetic moment. Thus, the electron Hamiltonian is
\begin{equation}
\label{eq:HB}
H_{el}=-\frac{\hbar^2}{2m}\partial_r^2+\mu_e\bm{B}_{Ov}\cdot\bm{\sigma}.
\end{equation}
In Eq. (\ref{eq:HB}) we do not include electron-electron interactions explicitly. In the following to evaluate the internal field $\bm{B}_{i}^N$ we use Eqs. (C4), and (C5) from Ref. \onlinecite{meng:arxiv14}. In these equations electron-electron interaction is significant (for example, for the critical temperature of the helical polarization \cite{braunecker:prb09, braunecker:prl09, meng:arxiv14, stano:prb14}) and therefore is included.

To describe the nuclear polarization in the wire when a bias voltage is applied, we will first investigate the behaviour of one nuclear spin placed in an effective field of all others, Eq. (\ref{eq:RKKYAveraged}), and interacting with the bath of electrons described by Eq. (\ref{eq:HB}).

\section{Helical electrons and finite voltage}
\label{sec:BathAsOneDimensionalChiralElectrons}
To find how a nuclear spin is affected by the electrons when the bias voltage is applied, we first consider  the properties of the electron bath in the wire.
As already mentioned in Sec. \ref{sec:TheModel}, the electrons are moving in the  Overhauser field  
produced by the nuclear spins [see Eqs. (\ref{eq:BOv}), (\ref{eq:HB})].
As the nuclear spins form a helix in equilibrium, this particular Overhauser field, denoted by $\bm{B}_h$, is also helical.
Consequently, the electron spectrum is
\begin{equation}
\varepsilon_\pm=\frac{\hbar^2(k^2+k_F^2)}{2m}\pm\frac{1}{m}\sqrt{m^2\mu_e^2B_h^2+\hbar^4k^2 k_F^2},
\end{equation}
where $k$ is the electron wavevector, and $\varepsilon_-$ and $\varepsilon_+$ denote the lower and upper subbands respectively. They are split by the gap $2\mu_e B_h$ at $k=0$. The corresponding wavefunctions are
\begin{eqnarray}
\label{eq:WavefunctionFull}
\Psi_{k,-}(r)&=&\frac{e^{ikr}}{\sqrt{L}}[e^{-ik_Fr}\cos{\frac{\theta_k}{2}}\ket{\uparrow}+e^{ik_Fr}\sin{\frac{\theta_k}{2}}\ket{\downarrow}], \mbox{\ \ \ \ }\\
\label{eq:WavefunctionFull2}
\Psi_{k,+}(r)&=&\frac{e^{ikr}}{\sqrt{L}}[e^{ik_Fr}\cos{\frac{\theta_k}{2}}\ket{\downarrow}-e^{-ik_Fr}\sin{\frac{\theta_k}{2}}\ket{\uparrow}], \mbox{\ \ \ \ }
\end{eqnarray}
where $\cos{\theta_k}=\frac{\hbar^2 k k_F}{\sqrt{(\hbar^2 k k_F)^2+(m\mu_e B_h)^2}}$ and $\sin{\theta_k}=\frac{-m\mu_e B_h}{\sqrt{(\hbar^2k k_F)^2+(m\mu_e B_h)^2}}$, and $\ket{\uparrow}$, $\ket{\downarrow}$ denote the spin states with spin up and spin down respectively, where $\bm{u}$ sets the quantization axis.
These expressions of the wavefunctions can be simplified since typically the ratio $\Lambda\equiv \mu_e B_{max}/\varepsilon_F\ll 1$, where $B_{max}$ is the maximum Overhauser field when all nuclei are fully polarized along a given direction. 
For example, for a GaAs quantum wire $\mu_e B_{max}\simeq 67\ \mu\mbox{eV}$, while $\varepsilon_F\simeq10 \mbox{ meV}$, which gives $\Lambda\simeq0.0067$. Consequently, we can use $\Lambda$ as a small parameter.  

We expand Eq. (\ref{eq:WavefunctionFull}) in leading order of $\Lambda$ and for the states within the partial gap   we get 
\begin{equation}
\label{eq:SimplifiedWavefunction}
\Psi_{k,-}(r)\approx \frac{1}{\sqrt{L}}\left \{ \begin{matrix}
-e^{i(k-k_F)r}\ket{\xi_R},  \,\, k>0,   \\
\, \, \, \, e^{i(k+k_F)r} \ket{\xi_L}, \,\, k<0,  \\ 
  \end{matrix} \right. 
\end{equation}
where for right-moving electrons $(k>0)$ the spinor is $\ket{\xi_R}=\ket{\uparrow}$, and for left-moving $(k<0)$ it is $\ket{\xi_L}=\ket{\downarrow}$.  Therefore, within our approximation the electronic states in the partial gap are helical: the spin is determined by the propagation direction, and is opposite for left-moving and right-moving electrons. 

\begin{figure}[tb]
\begin{center}
\includegraphics[width=\linewidth]{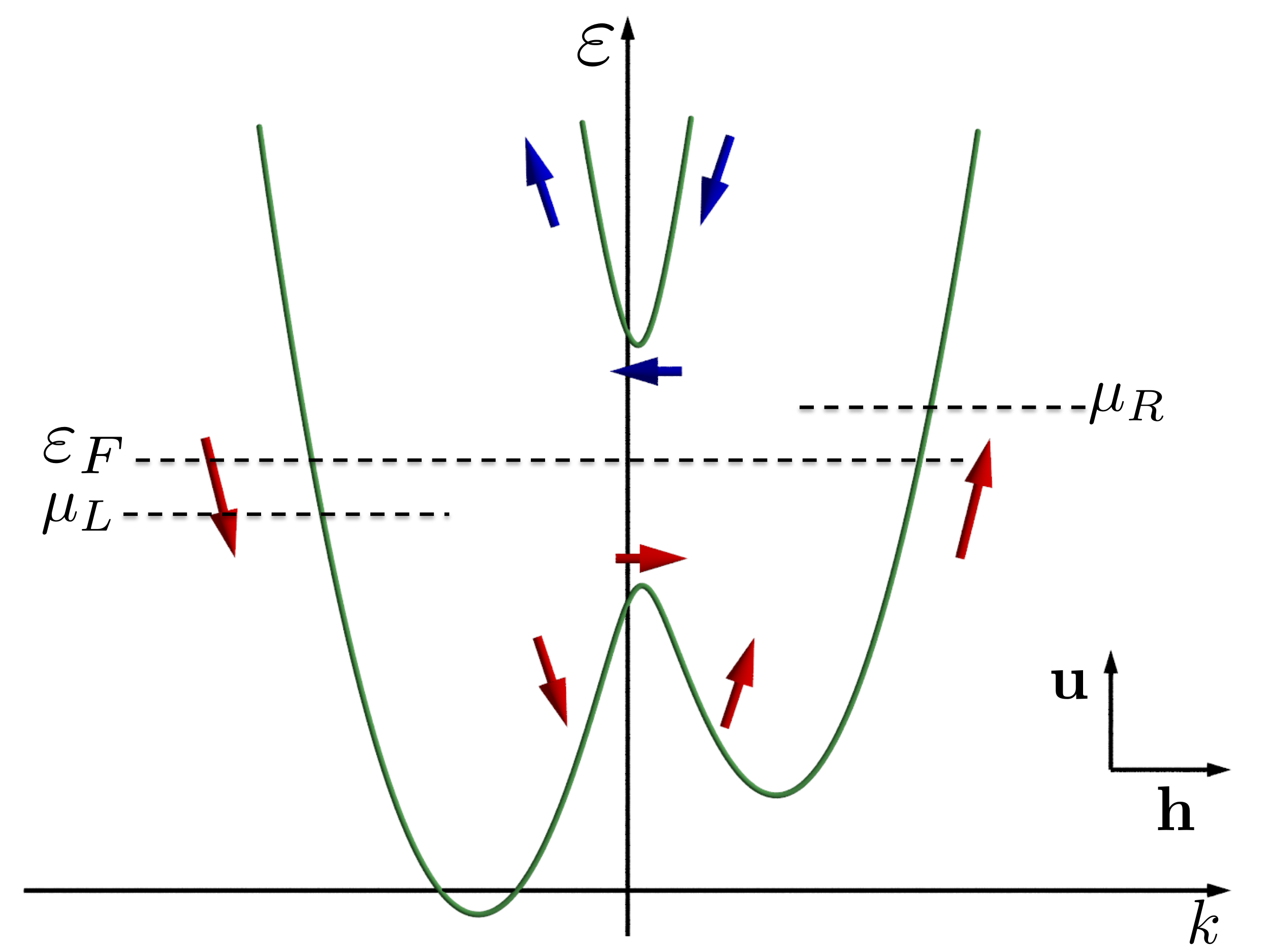}
\caption{Sketch of the energy spectrum given in Eq.  (\ref{eq:Spectrum}) and the direction of the electron spins in the presence of the helical Overhauser field $\bm{B_h}$ and the uniform Overhauser field $\bm{B_u}$ perpendicular to the plane of the helix. Red arrows denote the spin directions of the electrons in the lower subband, and the blue arrows label the spin directions for the upper subband. The coordinate system for the spins is formed by $\bm{h}$ and $\bm{u}$ shown in the right lower corner. The chemical potentials for left- and right-movers are denoted as $\mu_L$ and $\mu_R$, respectively. The voltage applied to the wire is $eV_{RL}=\mu_R-\mu_L$.}
\label{fig:spectrum}
\end{center}
\end{figure}

Next, we consider the voltage applied to the wire and define it as the difference between the chemical potentials for the left- and right-moving electrons (see Fig. \ref{fig:spectrum}). Assuming a ballistic wire, the chemical potential of a given branch is constant in space. Concerning the nuclear spins, only the electron states from the partial gap are relevant, because we consider the case of small voltage, i.e. $eV<2\mu_e B_h$, where $-e$ is the electron charge, and small temperatures $T$, {\it i.e.}, $k_BT<2\mu_eB_h$. This means that we adopt two approximations. First, we neglect the influence from electron states which are not in the partial gap (the upper ($+$) subband  is neglected completely), because their contribution to transport is exponentially small, proportional to $\mbox{exp}[(-\mu_eB_h+eV/2)/k_BT]$. Second, we use Eq. (\ref{eq:SimplifiedWavefunction}) for the electron wavefunctions, which means that we consider it in leading order of $k_BT/\varepsilon_F$, $eV/\varepsilon_F$, and $\Lambda$.  Therefore, 
for a description of the electron system in terms of  a heat bath that causes the relaxation of the nuclear spins,
we take into account two branches: left- and right-moving electrons with spins $\ket{\downarrow}$ and $\ket{\uparrow}$, respectively. 

With the polarity as assumed in Fig. \ref{fig:spectrum}, the applied voltage depletes the left (L) branch and increases the population of the right (R) branch. This imbalance in population opens up an additional phase space for the electrons to backscatter - predominantly from R to L. Because of the helical character of the states, such backscattering is accompanied by an electron spin flip (from $\ket{\uparrow}$ to $\ket{\downarrow}$). This, in turn, is enabled by the total spin-conserving hyperfine interaction Eq. (\ref{eq:HyperfineHamiltonian}), so that each electron spin flip is compensated by a nuclear spin flip in the opposite direction. In this way a {\it uniform} nuclear polarization along the $\bm{u}$ direction is built up. We denote $\bm{B}_u$ as the Overhauser field corresponding to this uniform polarization.

The spectrum of electrons moving in the total Ovehauser field 
$\bm{B}_h+\bm{B}_u$
reads
\begin{equation}
{\varepsilon}_{u,\pm}=\frac{\hbar^2(k^2+k_F^2)}{2m}\pm\sqrt{\mu_e^2B_h^2+\left[\frac{\hbar^2k k_F}{m}-\mu_e B_u\right]^2},
\label{eq:Spectrum}
\end{equation}
see Fig. \ref{fig:spectrum}. The asymmetry of the spectrum is due to the uniform Overhauser field $\bm{B}_u$. The corrections to the wave functions, Eq. (\ref{eq:SimplifiedWavefunction}), due to $\bm{B}_u$  are negligible in leading order of $\Lambda$. 

We note that from Eqs. (\ref{eq:WavefunctionFull}) and (\ref{eq:WavefunctionFull2}) it follows that the electron spins  become also polarized, 
thereby producing a Knight  shift  acting as an effective magnetic field $\bm{B}_{j}^e$ back on the nuclear spins. This Knight shift is defined as $\langle H_{hyp}\rangle_{el}=\mu_N \bm{B}_{j}^e\cdot \bm{I}_j$, where $\langle ...\rangle_{el}$ denotes averaging over the eigenstates of the Hamiltonian in Eq. (\ref{eq:HB}) with populations defined by the voltage. In this work, however, we can neglect $\bm{B}_{j}^e$ with respect to $\bm{B}_{j}^N$ produced by the RKKY interaction.\cite{braunecker:prb09,meng:arxiv14}

\section{Bloch Equation for the total nuclear spin in the wire}
\label{sec:BlochEquation}
To investigate the time-dynamics of the nuclear spins, we apply the standard Bloch-Redfield theory to our problem, which is valid for weak coupling between spin system and bath degrees of freedom,\cite{blum:book,golovach:prl04} as is the case here. First, we write down the Bloch equation for the average $\langle \bm{I}_j\rangle$ of the $j$th nuclear spin. By applying Eqs. (7)-(11) from Ref. \onlinecite{golovach:prl04} to our  Eqs. (\ref{eq:HyperfineHamiltonian}), (\ref{eq:RKKYAveraged}), and (\ref{eq:HB}), we get (for more details see Appendix \ref{app:BlochEquation})
\begin{equation}
\partial_t\langle \bm{I}_j\rangle=\bm{\omega}_j\times\langle\bm{I}_j\rangle-\bm{\Gamma}_j\langle\bm{I}_j\rangle+\bm{\Upsilon}_j, 
\label{eq:BlochEquationi}
\end{equation}
where  $\bm{\omega}_j=\mu_N\bm{B}_{j}^N/\hbar$ determines the precession, the relaxation tensor $\bm{\Gamma}_j$ the decay and the inhomogeneous vector term $ \bm{\Upsilon}_j$ the stationary value of $ \langle \bm{I}_j\rangle$. Both, $\bm{\Gamma}_j$ and $ \bm{\Upsilon}_j$ 
are expressed in terms of time correlators (see App. A)
\begin{equation}
\mathcal{J}_{nl}(\omega)=\frac{1}{2\hbar^2}\int_0^\infty e^{-i\omega t}\langle\delta B_n(0) \delta B_l(t)\rangle_{el} dt,
\label{eq:Correlator}
\end{equation}
where $t$ is time, the indexes $n,\ l$ label the components of  the effective fluctuating internal field $\delta\bm{ B}$ defined via $H_{hyp}-\langle H_{hyp}\rangle_{el}=\delta\bm{ B}\cdot\bm{I}_j$. The time-dependence follows from the interaction representation $\delta\bm{ B}(t)=e^{iH_{el}t/\hbar}\delta\bm{ B} e^{-iH_{el}t/\hbar}$. 
We note that above equations are valid for a spin 1/2. However, it is well-known~\cite{slichter:book} that the relaxation time of a spin into its stationary value
does not depend on the spin length (in Born approximation). Thus, we will assume that our results apply for arbitrary spins.

As follows from Secs. \ref{sec:TheModel} and \ref{sec:BathAsOneDimensionalChiralElectrons}, we can define the expectation value of a nuclear spin at position $R=0$ as
\begin{equation}
\langle\bm{I}_0\rangle/I=p_h\bm{h}+p_u\bm{u}, 
\end{equation}
where $0 \leq p_{h,u} \leq 1$ denote the polarizations along the two orthogonal directions $\bm{h}$ and $\bm{u}$, respectively.

We also introduce position-independent tensors $\bm{\Gamma}_0$ and $\bm{\Upsilon}_0$ in the rotated frame defined by the rotation matrix $\mathcal{R}_{\bm{u}, 2k_FR_j}^\dagger$, via
\begin{eqnarray}
\label{eq:Gamma0}
\bm{\Gamma}_j&=&\mathcal{R}_{\bm{u}, 2k_FR_j} \bm{\Gamma}_0 \mathcal{R}^\dagger_{\bm{u}, 2k_FR_j}, \\
\label{eq:Upsilon0}
\bm{\Upsilon}_j&=&\mathcal{R}_{\bm{u}, 2k_FR_j}\bm{\Upsilon}_0.
\end{eqnarray} 
Having Eqs. (\ref{eq:omega0}), (\ref{eq:BlochEquationi}), (\ref{eq:Gamma0}), and (\ref{eq:Upsilon0}), we can describe the time-evolution of the nuclear spin $\bm{I}_j$ in the rotated frame.

Eventually we are interested in the dynamics of the total (macroscopic) polarizations, rather than the one of an individual nuclear spin. We therefore introduce the total nuclear spin in the rotated frame $\sum_j \mathcal{R}^\dagger_{\bm{u}, 2k_FR_j}\langle \bm{I}_j\rangle\equiv NN_\perp\langle\bm{I}_0\rangle$, and write the equation of motion for it using Eqs. (\ref{eq:omega0}), (\ref{eq:BlochEquationi}), (\ref{eq:Gamma0}), and (\ref{eq:Upsilon0}). We get
\begin{equation}
\label{eq:BlochEquationTotal}
\partial_t \langle\bm{I}_0\rangle=-\Omega (\langle\bm{I}_0\rangle\cdot\bm{h})\bm{h}\times \bm{u}-\bm{\Gamma}_0\langle\bm{I}_0\rangle+\bm{\Upsilon}_0,
\end{equation}
where we denoted $\Omega=Ip_u|J_{2k_F}|/\hbar$. The first term implies a rotation of the helical direction $\bm{h}$, around the axis $\bm{u}$ with frequency $\Omega$. This can be seen by introducing a time-dependent vector $\bm{h}(t)=\mathcal{R}_{\bm{u},\alpha(t)}\bm{h}$, where $\alpha(t)=\int_0^t\Omega d\tau$. In the Born-Oppenheimer approximation, the tensors $\bm{\Gamma}_j$ and $\bm{\Upsilon}_j$ are functions of the instantaneous values of $\bm{h}$ and $\bm{u}$, so we write
\begin{eqnarray}
\bm{\tilde{\Gamma}}_j&=&\mathcal{R}_{\bm{u},\alpha(t)}\bm{\Gamma}_j \mathcal{R}^\dagger_{\bm{u},\alpha(t)}, \\
\bm{\tilde{\Upsilon}}_j&=&\mathcal{R}_{\bm{u},\alpha(t)}\bm{\Upsilon}_j.
\end{eqnarray}
With this the time evolution of $\langle\bm{I}_0\rangle$ in the rotating frame, $\langle\bm{\tilde{I}}_0\rangle=\mathcal{R}_{\bm{u},\alpha(t)}\langle\bm{I}_0\rangle$, is described by
\begin{equation}
\partial_t\langle\bm{\tilde{I}}_0\rangle=-\bm{\tilde{\Gamma}}_0\langle\bm{\tilde{I}}_0\rangle+\bm{\tilde{\Upsilon}}_0.
\label{eq:BlochEquationFinal}
\end{equation}
Using this equation and properties of the electron bath discussed in Sec. \ref{sec:BathAsOneDimensionalChiralElectrons}, we can describe the polarization of the nuclear spins in the wire as function of temperature and voltage.

\section{Resulting Polarizations}
\label{sec:Polarization}
To find the polarizations $p_h$ and $p_u$ from Eq. (\ref{eq:BlochEquationFinal}) we now evaluate the tensors $\bm{\Gamma}_0$ and $\bm{\Upsilon}_0$ explicitly. For that we first evaluate the correlator $\mathcal{J}_{nl}(\omega)$. Using Eqs. (\ref{eq:SimplifiedWavefunction}) and (\ref{eq:Spectrum}) we get
\begin{eqnarray}
\label{eq:CorrelatorElectrons}
&&\mathcal{J}_{nl}(\omega)=\frac{A^2a^2}{32\hbar^3\pi v_F^2N_\perp^2}\sum_{a,b\in \{L,R\}} M_{nl}^{ab}Q_{ab},\\ 
&&M_{nl}^{ab}=\langle\xi_a|\sigma_n|\xi_b\rangle\langle\xi_b|\sigma_l|\xi_a\rangle, \\ 
&&Q_{ab}=\int d\varepsilon f\left(\varepsilon+eV_{ba}/2\right)\left[1-f\left(\varepsilon+\hbar\omega-eV_{ba}/2\right)\right], \mbox{ \ \ \ \  }
\label{eq:Iab}
\end{eqnarray}
where $eV_{ba}=\mu_b-\mu_a$ is the difference between chemical potentials of branch $b$ and $a$, with $a$ and $b$ denoting $L$ (left-movers) or $R$ (right-movers). Here we also use the Fermi distribution function $f(\varepsilon)=\left[\mbox{exp}\left[\varepsilon/(k_BT)\right]+1\right]^{-1}$. As was mentioned in Sec. \ref{sec:BathAsOneDimensionalChiralElectrons}, we consider voltages and temperatures smaller than the partial gap $2\mu_e B_h$ given by the helical polarization. Therefore the term $f(\varepsilon+eV_{ba}/2)[1-f(\varepsilon+\hbar\omega-eV_{ba}/2)]$ allows us to consider only the energy window of $\pm \mu_e B_h$ around $\varepsilon_F$, because $f(\varepsilon)$ decays exponentially for $\varepsilon/k_BT\gg1$. Consequently, we approximate the electron density of states (per spin) by $\nu(\varepsilon)\approx\nu(\varepsilon_F)$. Up to first order in $\Lambda$, we have $\nu(\varepsilon_F)=1/(\pi\hbar v_F)$, where $v_F=\varepsilon_F/(\hbar k_F)$ is the Fermi velocity of the electrons.

Having obtained $\mathcal{J}_{nl}(\omega)$, it is straightforward to calculate  $\bm{\Gamma}_0$ and $\bm{\Upsilon}_0$, using Eqs. (\ref{eq:Gammad})-(\ref{eq:Upsilon}) and (\ref{eq:CorrelatorElectrons})-(\ref{eq:Iab}). We can then solve Eq. (\ref{eq:BlochEquationFinal}) for the steady state polarizations (keeping $\omega_0$ as a constant) and obtain
\begin{widetext}
\begin{eqnarray}
p_h&=&\frac{4\hbar\omega_0}{(\hbar\omega_0-eV)\coth\left(\frac{\hbar\omega_0-eV}{2k_BT}\right)+(\hbar\omega_0+eV)\coth\left(\frac{\hbar\omega_0+eV}{2k_BT}\right)+2\hbar\omega_0\coth\left(\frac{\hbar\omega_0}{2k_BT}\right)},
\label{polarization_ph}\\
p_u&=&\frac{4\hbar\omega_0\frac{(\hbar\omega_0-eV)\coth\left(\frac{\hbar\omega_0-eV}{2k_BT}\right)-(\hbar\omega_0+eV)\coth\left(\frac{\hbar\omega_0+eV}{2k_BT}\right)}{(\hbar\omega_0-eV)\coth\left(\frac{\hbar\omega_0-eV}{2k_BT}\right)+(\hbar\omega_0+eV)\coth\left(\frac{\hbar\omega_0+eV}{2k_BT}\right)+2\hbar\omega_0\coth\left(\frac{\hbar\omega_0}{2k_BT}\right)}+4eV}{(\hbar\omega_0-eV)\coth\left(\frac{\hbar\omega_0-eV}{2k_BT}\right)+(\hbar\omega_0+eV)\coth\left(\frac{\hbar\omega_0+eV}{2k_BT}\right)+2eV\coth\left(\frac{eV}{2k_BT}\right)}.
\label{polarization_pu}
\end{eqnarray}
\end{widetext}

However, from Eq. (\ref{eq:omega0}) it follows that $\hbar\bm{\omega}_0=2p_hN_\perp IJ_{2k_F}\bm{h}$, {\it i.e.}, $\omega_0$ depends on $p_h$. This leads to non-linear algebraic equations for two unknowns, $p_u$ and $p_h$, which we solve numerically  using material parameters for GaAs (analytical expressions for small deviations of the polarizations are given below).
We  plot the values obtained in this way and discuss their behaviour as a function of voltage and temperature, the  experimental parameters that are most directly accessible.

\begin{figure}[tb]
\begin{center}
\includegraphics[width=\linewidth]{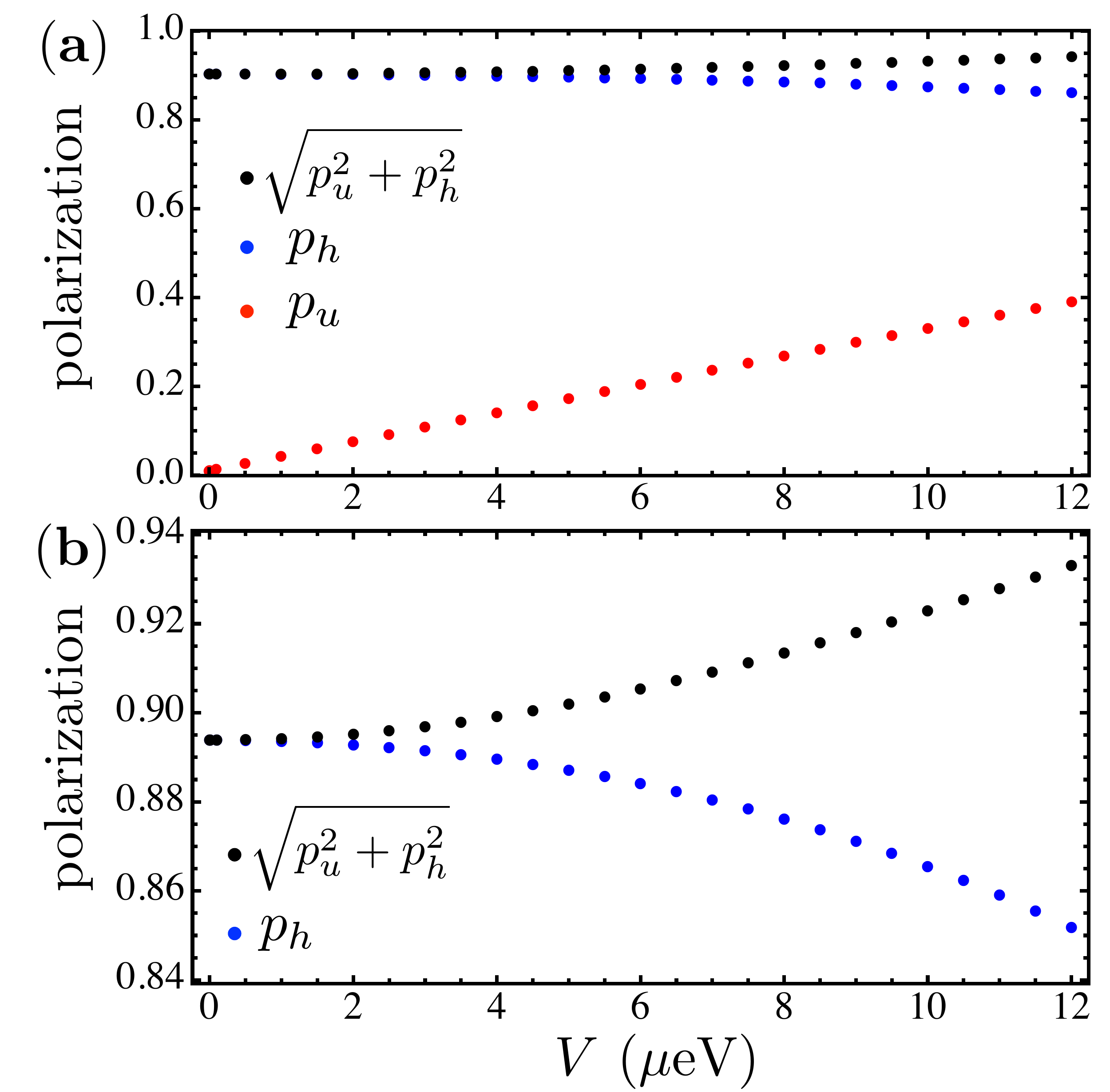}
\caption{(a) The voltage dependence of the polarization $p_h$ along helical direction $\bm{h}$ (blue), polarization $p_u$ in the direction of $\bm{u}$ perpendicular to the helix plane (red), and the overall polarization of nuclei $\sqrt{p_u^2+p_h^2}$ (black). (b) Enlarged from (a) the voltage dependence of $p_h$ and $\sqrt{p_u^2+p_h^2}$. We use $T=90\mbox{ mK}$ and other parameters as given in the text.}
\label{fig:PolarizationVoltage}
\end{center}
\end{figure}

The voltage dependence of the polarizations is shown in Fig. \ref{fig:PolarizationVoltage}. We can see that the polarization $p_u$ grows faster with voltage than $p_h$ decays, therefore the overall polarization of the nuclei $\sqrt{p_u^2+p_h^2}$ grows with voltage, too. This means that the nuclear spins are more polarized  when a voltage is applied than when they are in equilibrium at the same temperature. We also note that having a non-zero component $p_u$ means that nuclear spins have a conical polarization, rather than a helical one.
To plot Fig. \ref{fig:PolarizationVoltage} we used Eqs. (C4), and (C5) from Ref. \onlinecite{meng:arxiv14} as was mentioned above, where the dependence of $\hbar\omega_0$ on temperature is described in detail. To evaluate $\hbar\omega_0$ we used the characteristic values for GaAs: the Fermi velocity $v_F=2.3\times 10^5\mbox{ m/s}$, and the number of nuclei in the wire cross-section $N_\perp=1300$. For the expression for $J_{2k_F}$ taken from Ref. \onlinecite{meng:arxiv14} we use the electron-electron interaction Luttinger liquid parameter $K_\rho=0.2$ and the absolute value of spin $I=3/2$. For the constants described above and at $T=90\mbox{ mK}$ and $p_u=0.1$ the rotation frequency of the nuclear spin helix is $\Omega\approx1.5\times 10^6\mbox{ s}^{-1}$. 

\begin{figure}[tb]
\begin{center}
\includegraphics[width=\linewidth]{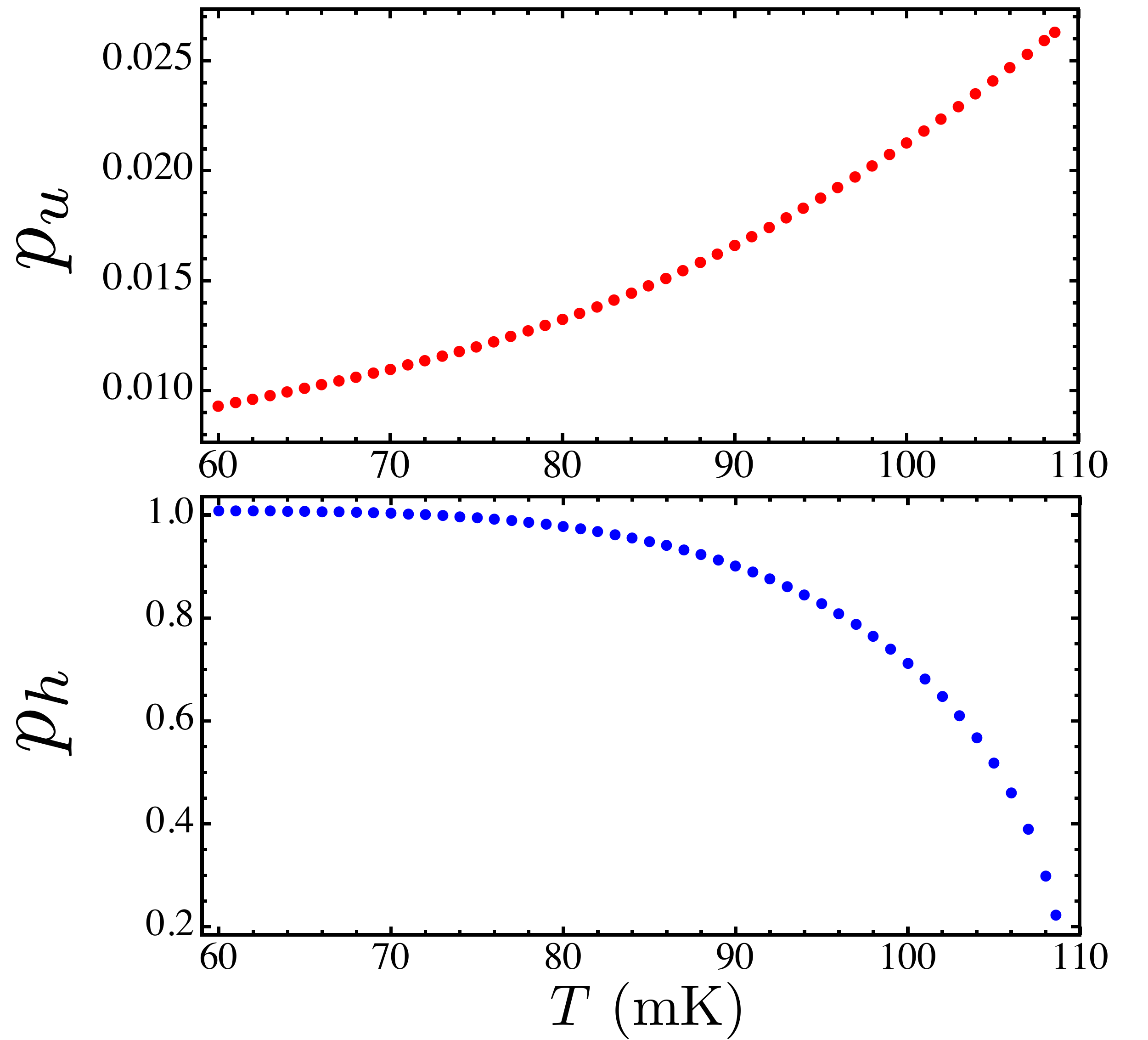}
\caption{
Plot of the temperature dependence of the polarization  $p_u$ 
(upper panel, red) and the polarization
$p_h$  
(lower panel, blue).
For these plots the same parameters were used as in Fig. \ref{fig:PolarizationVoltage} and the applied voltage is $eV=0.5\ \mu\mbox{eV}$. We note that our calculation is valid for $eV, k_BT<2\mu_eB_h$, therefore the smallest value of $p_h$ we consider here is $p_h\simeq0.2$. }
\label{fig:PolarizationTemperature}
\end{center}
\end{figure}

It is natural to expect that high temperature destroys the nuclear helical order.\cite{braunecker:prl09, braunecker:prb09, stano:prb14, meng:arxiv14} Indeed, Fig. \ref{fig:PolarizationTemperature} shows that the helical polarization $p_h$ decays with temperature and then drops in magnitude around $T\simeq 109\mbox{ mK}$. As our calculation is valid for $eV, k_BT<2\mu_eB_h$, the smallest value of $p_h$ allowed by self-consistency for our parameters is $p_h\simeq0.2$. From Fig. \ref{fig:PolarizationTemperature} it also follows that the polarization $p_u$ grows with temperature. This growth is explained by the fact that due to higher temperature the electron states with higher energy become occupied. This makes the nuclear spin flip more probable. It is obvious that there is a temperature where the polarization $p_u$ gets destroyed, however, for the  range of temperatures given in Fig.  \ref{fig:PolarizationTemperature} $p_u$ does grow, whereas the helical polarization $p_h$ decays significantly. The decay of $p_h$ with temperature is rapid, while the growth of $p_u$ is less pronounced. Therefore, the overall nuclear polarization in the wire strongly decays with increasing temperature. 
For the parameters we used for Fig. \ref{fig:PolarizationTemperature} the effect of temperature on $p_h$ is stronger than the one of a finite voltage. The initial temperature scaling of $p_h$ away from unity
(see Fig. \ref{fig:PolarizationTemperature}) can be obtained readily from Eq.~(\ref{polarization_ph}) by treating  $1-p_h$ as a small perturbation. This yields
\begin{eqnarray}
p_h&\approx &1-\frac{2}{1+e^{\frac{\kappa}{T^{g}}}} 
\cdot
\frac{1}{1-\frac{\kappa}{2T^{g}}\mbox{ sech}^2{\left(\frac{\kappa}{2T^{g}}\right)}}\\
&\approx & 1-2e^{-\frac{\kappa}{T^{g}}},
\end{eqnarray}
where the first equality holds well for the temperature interval $60 \mbox{mK}<T<90 \mbox{mK}$, while the second one is a good approximation for $60 \mbox{mK}<T<80 \mbox{mK}$. 
Here we denoted 
$g=3-\frac{4K_\rho}{\sqrt{2(1+K_\rho^2)}}$, and the temperature-independent parameter $\kappa=2N_\perp I|J_{2k_F}|T^{g-1}/k_B $ depends on the material and geometrical properties of the sample (see Eq. (11) and  Eqs. (C4) and (C5) of Ref. \onlinecite{meng:arxiv14}). 
For  $K_\rho=0.2$ (chosen for the plots) we get $g=2.4$ (we recall that
$K_\rho=g=1$ corresponds to vanishing electron-electron interactions).

The initial decrease of $p_h$ due to voltage in Fig. \ref{fig:PolarizationVoltage} for $V<3\ \mu$eV scales as
\begin{equation}
p_h\approx \alpha-\gamma V^2,
\end{equation}
where $\alpha$ and $\gamma$ depend on material and geometrical parameters of the nanowire and on temperature.

Finally, we mention that recent progress in nuclear spin magnetometry on nanowires\cite{peddibhotla:nature13} has opened the perspective to measure the
nuclear spin polarizations directly and thus to test the predictions made here.
Moreover, due to the helical nuclear polarization which acts on electrons as an Overhauser field $\bm{B}_h$ there is a partial gap in the electron spectrum [see Eq. (\ref{eq:Spectrum})]. As a result, the conductance of a ballistic nanowire is less than $2e^2/h$ for sufficiently low temperatures and $V<2\mu_eB_h$.~\cite{braunecker:prl09, braunecker:prb09,scheller:prl14,meng:arxiv14}
As was shown above, the polarization $p_h$, and consequently $\bm{B}_h$, decrease with increasing voltage and temperature. 
We thus expect qualitatively that the conductance of the wire will increase with the decrease of the partial gap $2\mu_eB_h \propto p_h$.~\cite{footnote_conductance}

\section{Conclusions}
\label{sec:Conclusions}
We have shown that due to the hyperfine interaction between electrons and nuclei in the wire the applied voltage changes the form of the nuclear polarization and its amplitude. Assuming that in equilibrium there is a helical nuclear polarization $p_h$ present in the wire due to RKKY interaction, a bias voltage induces a uniform polarization $p_u$ perpendicular to the helix plane. Due to this polarization the nuclear spin helix starts to rotate around the axis perpendicular to the helical plane. 
When a non-zero polarization $p_u$ buidls up, the nuclear polarization changes from helical to conical.

We have also presented the voltage dependence of $p_u$ and $p_h$ and seen that $p_u$ increases with voltage, whereas $p_h$ decreases. Following from these two effects the overall nuclear polarization in the wire grows with voltage. Remarkably, $p_u$ grows with temperature in the considered range of temperatures. This is because the nuclear spin flip becomes more probable as electrons occupy higher energy states. This thermal activation effect is noticeable for the considered regime $\hbar\omega_0> eV$. 
The growth of the overall polarization $\sqrt{p_u^2+p_h^2}$ with voltage and the growth of $p_u$ with temperature are intriguing and {\it a priori} non-obvious effects. The polarization effects predicted here might be observed in transport experiments~\cite{scheller:prl14} or more directly via cantilever based nanoscale magnetometry.~\cite{peddibhotla:nature13}

\begin{acknowledgments} 
We thank V. N. Golovach, D. Becker, B. Braunecker, L. Glazman, C. Kloeffel, T. Meng, and C. Orth for helpful discussions and acknowledge support from the Swiss NF, NCCR QSIT, and S$^3$NANO.   
\end{acknowledgments}


\appendix

\section{Bloch equation for one nuclear spin}
\label{app:BlochEquation}
To write down the Bloch equation for the total nuclear spin in the wire, we use Eqs. (7)-(11) from Ref. \onlinecite{golovach:prl04}. Here we present them adopted to our case of a nuclear spin interacting with the bath of electrons and placed into the effective field produced by all other nuclear spins in the wire.

The Bloch equation for the $n$th nuclear spin reads
\begin{equation}
\label{eq:appBlochEquation}
\partial_t\langle \bm{I}_n\rangle=\bm{\omega}_n\times\langle\bm{I}_n\rangle-\bm{\Gamma}_n\langle\bm{I}_n\rangle+\bm{\Upsilon}_n.
\end{equation}
To express tensors $\bm{\Gamma}_n$ and $\bm{\Upsilon}_n$ we introduce a unit vector $\bm{l}$ along $\bm{\omega}_n$, i.e. $\bm{\omega}_n=\omega_n\bm{l}$. The tensor $\bm{\Gamma}_n$ consists of a dephasing part $\bm{\Gamma}^d_n$ which comes from energy conserving processes and a pure relaxation part $\bm{\Gamma}_n^r$, which comes from the energy exchange with the bath \cite{golovach:prl04, borhani:prb06} (played here by the electron system),
\begin{eqnarray}
\label{eq:Gammad}
&\bm{\Gamma}^d_{n,ij}&=[\delta_{ij}l_pl_q\mathcal{J}^+_{pq}(0)-l_il_p\mathcal{J}_{pj}^+(0)], \\ 
&\bm{\Gamma}^r_{n,ij}&=[\delta_{ij}(\delta_{pq}-l_pl_q)\mathcal{J}_{pq}^+(\omega_n)-\\ \nonumber &-&(\delta_{ip}-l_il_p)\mathcal{J}_{pj}^+(\omega_n)-\delta_{ij}\epsilon_{kpq}l_k\mathcal{I}_{pq}^-(\omega_n)+\\ \nonumber &+&\epsilon_{ipq}l_p\mathcal{I}^-_{qj}(\omega_n)].
\end{eqnarray}
Here, the indexes $i,\ j$ denote components of tensors, and we use the Einstein convention of summation over repeated indexes. Further, $\epsilon_{pqk}$ is the Levi-Civita symbol and $\delta_{ij}$ the Kronecker delta, while $l_k$ denotes the $k$th component of vector $\bm{l}$. The inhomogeneous part of the Bloch equation $\bm{\Upsilon}_n$ reads \cite{golovach:prl04, borhani:prb06}
\begin{eqnarray}
\label{eq:Upsilon}
\bm{\Upsilon}_{n,i}=\frac{1}{2}(l_j\mathcal{J}_{ij}^-(\omega_n)-l_i\mathcal{J}_{jj}^-(\omega_n)+\epsilon_{ipq}\mathcal{I}_{pq}^+(\omega_n)+\mbox{ \ \ \ } \\ \nonumber+\epsilon_{iqk}l_kl_p[\mathcal{I}^+_{pq}(\omega_n)-\mathcal{I}_{pq}^+(0)]),\mbox{\ \ \ }
\end{eqnarray} 
where $i$ denotes the component of $\bm{\Upsilon}_n$. The terms $\mathcal{J}_{ij}^\pm(\omega),$ $\mathcal{I}^\pm_{ij}(\omega)$ are defined as
\begin{eqnarray}
\mathcal{J}_{ij}^\pm(\omega)&=&\mbox{Re}[\mathcal{J}_{ij}(\omega)\pm \mathcal{J}_{ij}(-\omega)], \\ 
\mathcal{I}_{ij}^\pm(\omega)&=&\mbox{Im}[\mathcal{J}_{ij}(\omega)\pm \mathcal{J}_{ij}(-\omega)].
\end{eqnarray}
The term $\mathcal{J}_{ij}(\omega)$ is the Laplace transformation of the correlator of the fluctuating fields $\delta \bm{ B}$ at different times,
\begin{equation}
\mathcal{J}_{ij}(\omega)=\frac{1}{2\hbar^2}\int_0^\infty e^{-i\omega t}\langle\delta B_i(0) \delta B_j(t)\rangle_{el} dt,
\end{equation}
where $\delta \bm{ B}(t)=e^{iH_{el}t/\hbar}\delta\bm{ B} e^{-iH_{el}t/\hbar}$. 
Using Eq. (\ref{eq:appBlochEquation}) we expressed the Bloch equation for the total nuclear spin in the wire resulting in Eq. (\ref{eq:BlochEquationTotal}).


\begin{thebibliography}{}
\bibitem{footnote_conductance}
However, we caution that the  behaviour of the polarizations $p_{h,u}$ can only indicate  the trend for the conductance but does not give its precise dependence on temperature and voltage. 
Indeed, in transport new temperature and voltage effects emerge, especially 
 in the transition region where the partial gap is comparable to temperature and voltage.
\end{thebibliography}
\end{document}